\documentclass[prl,twocolumn,aps,showpacs,nofootinbib,floatfix,superscriptaddress]{revtex4-1}
\usepackage{amsmath}
\usepackage{amssymb}
\usepackage{graphicx}

\usepackage{txfonts}

\usepackage[colorlinks=true,linkcolor=blue,citecolor=blue,urlcolor=blue]{hyperref}

\newcommand{\be}{\begin{equation}}
\newcommand{\ee}{\end{equation}}
\newcommand{\beq}{\begin{eqnarray}}
\newcommand{\eeq}{\end{eqnarray}}
\newcommand{\bra}[1]{\ensuremath{\langle#1|}}
\newcommand{\ket}[1]{\ensuremath{|#1\rangle}}

\begin{document}

\title{Efficient entanglement criteria for discrete, continuous, and hybrid variables}
\author{Manuel Gessner}
\email{manuel.gessner@ino.it}
\affiliation{QSTAR, CNR-INO and LENS, Largo Enrico Fermi 2, I-50125 Firenze, Italy}
\affiliation{Istituto Nazionale di Ricerca Metrologica, Strada delle Cacce 91, I-10135 Torino, Italy}
\author{Luca Pezz\`e}
\affiliation{QSTAR, CNR-INO and LENS, Largo Enrico Fermi 2, I-50125 Firenze, Italy}
\author{Augusto Smerzi}
\affiliation{QSTAR, CNR-INO and LENS, Largo Enrico Fermi 2, I-50125 Firenze, Italy}
\date{\today}
%\pacs{}

\begin{abstract}
We provide a method to construct entanglement criteria for arbitrary multipartite systems of discrete or continuous variables and hybrid combinations of both. While any set of local operators generates a sufficient condition for entanglement of arbitrary quantum states, a suitable set leads to a necessary and sufficient criterion for pure states. The criteria are readily implementable with existing technology and reveal entanglement that remains undetected by the respective state-of-the-art methods for discrete and continuous variables.
\end{abstract}

\maketitle

\textit{Introduction.}---Determining whether an unknown quantum state is entangled or not is a highly complex and in general unsolved problem \cite{Braunstein,Horodecki,GuehneToth}. The fundamental role of entanglement in quantum physics renders this issue directly relevant for various fields ranging from quantum information to condensed-matter physics \cite{Horodecki,Fazio,NielsenChuang}. A large amount of theoretical methods to characterize entanglement has been proposed \cite{Braunstein,Horodecki,Mintert,GuehneToth,Huber,Guehne,Sperling,Gisin2,Dicke}, however only few of them can be formulated in terms of a feasible operational recipe, involving only a given set of accessible operators and a limited number of measurements. These are crucial requirements to render such methods relevant for scalable experimental applications in a broad range of scenarios. 

The development of methods for systems of either discrete or continuous variables has led to two distinct approaches, which are widely popular in present-day experiments \cite{Dicke,Chip,Strobel,Bohnet,Peise,Treps,Lucke,Schmied}. Continuous-variable systems, on the one hand, are typically analyzed with separability criteria involving uncertainty relations \cite{Duan,Simon,Giovannetti,Braunstein}. They provide a complete characterization of the entanglement of bipartite Gaussian states in terms of variances of suitably defined operators \cite{Duan,Simon}. These criteria have been further sharpened by means of entropic uncertainty relations \cite{Walborn} or moments of arbitrary order \cite{SV}, enabling them to detect a larger amount of non-Gaussian states, but their application remains limited to bipartite systems. 

On the other hand, a convenient method to detect entanglement in the discrete (e.g. multi-qubit) case is based on the violation of spin-squeezing inequalities \cite{Wineland92, Sorensen,TothPRA2009}. A larger class of entangled states, including the spin-squeezed ones, can be detected by the Fisher information \cite{PS09}, a fundamental quantity in estimation theory \cite{Helstrom}. However, as this method is specifically designed to detect only those states that lead to a metrological advantage -- only one of many applications of entangled states --, certain entangled states remain undetected, including some pure states \cite{Hyllus}.

In this article, we provide a unified approach to entanglement detection in arbitrary multipartite quantum systems, which proves to be more efficient than standard strategies for either discrete \cite{GuehneToth,PS09,Hyllus} or continuous variables \cite{Duan,Simon,Giovannetti,Walborn}. In particular, every pure entangled state can be detected. Furthermore, the method can be readily adapted to a wide range of present-day experimental setups, since any set of accessible local observables can be used to construct a separability criterion. This flexibility also opens up the possibility to detect hybrid entanglement between discrete and continuous variables, whose prospects as a platform for implementations of quantum information is currently being explored \cite{Bellini,Laurat,Andersen}.

Specifically, we show that all separable quantum states of $N$ parties must satisfy the following inequality:
\be \label{eq.sepbound}
F_{\hat{M}} \bigg[ \hat{\rho}_{\rm sep}, \sum_{i=1}^N \hat{A}_i \bigg] \leq 4\sum_{i=1}^N\mathrm{Var} \left( \hat{A}_i \right)_{\hat{\rho}_{\rm sep}}.
\ee
Here $\hat{A}_i$ is a local observable for the $i$th party, $\mathrm{Var}(\hat{A})_{\hat{\rho}}=\langle\hat{A}^2\rangle_{\hat{\rho}} -\langle\hat{A}\rangle_{\hat{\rho}}^2$ 
denotes the variance and the quantum expectation values are given as $\langle \hat{A}\rangle_{\hat{\rho}}=\mathrm{Tr}[\hat{A}\hat{\rho}]$. 
The quantity appearing on the left-hand side of Eq.~(\ref{eq.sepbound}) is the Fisher information \cite{FIfootnote,Helstrom}. It quantifies how sensitively changes of the parameter $\theta$ are detected when the initial state $\hat{\rho}$ is transformed by the unitary evolution $\hat{\rho}(\theta)=e^{-i\sum_j\hat{A}_j\theta} \hat{\rho} e^{i\sum_j\hat{A}_j\theta}$ and then observed by measurements of the observable $\hat{M}$ \cite{Helstrom,Giovannetti06,Paris,GiovannettiReview,Varenna}.
It is furthermore experimentally accessible \cite{Strobel}, see also \cite{Bohnet,Monroe,FIPhoton}, without any knowledge of the full quantum state \cite{PezzePRE,frowis,hauke}. The bound~(\ref{eq.sepbound}) holds for arbitrary observables $\hat{M}$, rendering the criterion robust against imperfect implementations of the measurement \cite{PezzePRE}.

Since Eq.~(\ref{eq.sepbound}) represents a necessary criterion for separability, its violation is a sufficient criterion for entanglement. The appearance of state-dependent variances on the right-hand side makes this criterion highly versatile since it holds for arbitrary local observables $\hat{A}_i$, independent of the Hilbert-space structure. %As exemplified below, these criteria allow to detect the entanglement of mixed continuous-variable states that cannot be detected with the strongest available uncertainty-based criteria \cite{Duan,Simon,Giovannetti,Walborn}. Moreover, we will also show that the entanglement of arbitrary pure states can be detected by violation of Eq.~(\ref{eq.sepbound}) for suitably chosen observables $\hat{A}_i$. 

Equation~(\ref{eq.sepbound}) expresses the trade-off for separable states $\hat{\rho}_{\rm sep}$ between the state's sensitivity quantified by the Fisher information and the variances of the local operators $\hat{A}_i$ generating the transformation. If quantum correlations between the parties are present, then the bound~(\ref{eq.sepbound}) can be violated. In fact, for arbitrary quantum states $\hat{\rho}$, a different bound, $F_{\hat{M}} [ \hat{\rho}, \sum_{i=1}^N \hat{A}_i ] \leq 4\sum_{i,j}\mathrm{Cov}(\hat{A}_i,\hat{A}_j)_{\hat{\rho}}$ holds, where $\mathrm{Cov}(\hat{A},\hat{B})_{\hat{\rho}}=\langle\hat{A}\hat{B}+\hat{B}\hat{A}\rangle_{\hat{\rho}} / 2 -\langle\hat{A}\rangle_{\hat{\rho}}\langle\hat{B}\rangle_{\hat{\rho}}$. The difference between this bound and Eq.~(\ref{eq.sepbound}) lies entirely in the absence of covariances between the subsystems $(i\neq j)$ in~(\ref{eq.sepbound}). 

To demonstrate Eq.~(\ref{eq.sepbound}) we first notice that the Fisher information can be maximized over all possible observables $\hat{M}$ \cite{BraunsteinPRL1994}: We have $F_{\hat{M}} \leq F_{Q}$, where $F_{Q} [ \hat{\rho}, \sum_{i=1}^N \hat{A}_i] = \max_{\hat{M}} F_{\hat{M}} [ \hat{\rho}, \sum_{i=1}^N \hat{A}_i]$ is a saturable bound (i.e. optimal observables can be constructed) called the quantum Fisher information \cite{Helstrom}. For an arbitrary separable state 
$\hat{\rho}_{\rm sep} = \sum_{\gamma}p_{\gamma} \ket{\varphi_\gamma} \bra{\varphi_\gamma}$, 
where $\ket{\varphi_\gamma} = |\varphi^{\gamma}_1\rangle\otimes\dots\otimes|\varphi^{\gamma}_N\rangle$
is a product state, $p_\gamma > 0$, $\sum_\gamma p_\gamma=1$ and $\ket{ \varphi^{\gamma}_i }$ is the state of the $i$th party, 
the chain of inequalities 
\begin{subequations} \label{eq.sepbound.c}
\begin{align} 
 F_Q\bigg[\hat{\rho}_{\mathrm{sep}},\sum_{i=1}^N \hat{A}_i\bigg]&\leq\sum_{\gamma}p_{\gamma}F_Q\bigg[\ket{\varphi_{\gamma}},\sum_{i=1}^N\hat{A}_i\bigg] \label{eq.proof1} \\
&=4\sum_{\gamma}p_{\gamma}\mathrm{Var}\bigg(\sum_{i=1}^N\hat{A}_i\bigg)_{\ket{\varphi_{\gamma}}} \label{eq.proof2} \\
&=4\sum_{\gamma}p_{\gamma} \sum_{i=1}^N \mathrm{Var}\big(\hat{A}_i\big)_{|\varphi_i^{\gamma}\rangle} \label{eq.proof3} \\
& \leq  4\sum_{i=1}^N \mathrm{Var}\big(\hat{A}_i\big)_{\hat{\rho}_{\mathrm{sep}}} \label{eq.proof4}
\end{align}
\end{subequations}
holds. In Eq.~(\ref{eq.proof1}) we used the convexity of the quantum Fisher information \cite{Varenna} and, in Eq.~(\ref{eq.proof2}), the general expression 
$F_Q\big[\ket{\psi},\hat{M}\big] = 4 (\Delta \hat{M})^2$, valid for pure states $|\psi\rangle$ and Hermitian operators $\hat{M}$ \cite{BraunsteinPRL1994}.
We have $\mathrm{Var}\big(\sum_{i=1}^N\hat{A}_i\big)_{\hat{\rho}}=\sum_{i,j=1}^N\mathrm{Cov}(\hat{A}_i,\hat{A}_j)_{\hat{\rho}}$ and $\mathrm{Cov}(\hat{A},\hat{A})_{\hat{\rho}} = \mathrm{Var}(\hat{A})_{\hat{\rho}}$. Equation~(\ref{eq.proof3}) is then obtained by noticing that
$\mathrm{Cov}(\hat{A}_i,\hat{A}_j)_{|\varphi_1^{\gamma}\rangle\otimes\dots\otimes|\varphi_N^{\gamma}\rangle}=0$ when $i\neq j$.
Therefore, $\mathrm{Var}\big(\sum_{i=1}^N\hat{A}_i\big)_{|\varphi_1^{\gamma}\rangle\otimes\dots\otimes|\varphi_N^{\gamma}\rangle} 
= \sum_{i=1}^N \mathrm{Var}\big(\hat{A}_i\big)_{|\varphi_i^{\gamma}\rangle}$. 
Finally, the last inequality (\ref{eq.proof4}) follows from the concavity of the variance, see also Ref.~\cite{Guehne04}. 

In the above derivation, no assumption is made about the local operators $\hat{A}_i$. 
In fact, any choice of $\hat{A}_i$ leads to a sufficient criterion for entanglement. However, %depending on the state $\hat{\rho}$, 
certain choices of operators may be better suited than others to detect the entanglement of a given state $\hat{\rho}$.
In order to construct the strongest possible criterion, we decompose each of the individual $\hat{A}_i$ in terms of an accessible set of operators $\hat{\mathbf{A}}_i=(\hat{A}_i^{(1)},\hat{A}_i^{(2)},\dots)^T$ in the Hilbert space $\mathcal{H}_{i}$ of the $i$th party ($i=1,...,N$). Thus, the local operators $\hat{A}_i$ are replaced by the expressions $\sum_{m=1}c_i^{(m)}\hat{A}_i^{(m)}=\mathbf{c}_i\cdot\hat{\mathbf{A}}_i$, where the $\mathbf{c}_i=(c^{(1)}_i,c^{(2)}_i,\dots)$ are vectors of coefficients. In this case, the full generator of the unitary transformation $\hat{A}(\mathbf{c})=\sum_{i=1}^N\mathbf{c}_i\cdot\hat{\mathbf{A}}_i$ is characterized by the combined vector $\mathbf{c}=(\mathbf{c}_1,\dots,\mathbf{c}_N)^T$. According to Eq.~(\ref{eq.sepbound}), the quantity
\be \label{eq.W}
W[\hat{\rho},\hat{A}(\mathbf{c})]=F_Q[\hat{\rho},\hat{A}(\mathbf{c})]-4\sum_{i=1}^N\mathrm{Var}(\mathbf{c}_i\cdot\hat{\mathbf{A}}_i)_{\hat{\rho}}
\ee
must be non-positive for arbitrary choices of $\mathbf{c}$ whenever the state $\hat{\rho}$ is separable. We can now maximize $W[\hat{\rho},\hat{A}(\mathbf{c})]$ by variation of $\mathbf{c}$ to obtain an optimized entanglement witness for the state $\hat{\rho}$, given the sets of available operators contained in $\mathcal{A}=\{\hat{\mathbf{A}}_1,\dots,\hat{\mathbf{A}}_N\}$.

To this aim let us first express the quantum Fisher information in matrix form as
$F_Q[\hat{\rho},\hat{A}(\mathbf{c})]=\mathbf{c}^TQ^{\mathcal{A}}_{\hat{\rho}}\mathbf{c}$,
where the spectral decomposition $\hat{\rho}=\sum_kp_k|\Psi_k\rangle\langle\Psi_k|$ defines $\left(Q^{\mathcal{A}}_{\hat{\rho}}\right)^{mn}_{ij}=2\sum_{k,l}\frac{(p_k-p_l)^2}{p_k+p_l}\langle\Psi_k|\hat{A}_i^{(m)}|\Psi_l\rangle\langle\Psi_l|\hat{A}_j^{(n)}|\Psi_k\rangle$ element-wise and the sum extends over all pairs with $p_k+p_l\neq 0$. The indices $i$ and $j$ refer to different parties (i.e. different Hilbert spaces), while the indices $m$ and $n$ label the respective local sets of observables within each Hilbert space. Similarly, for the list of operators $\mathcal{A}$, we can express the elements of the covariance matrix of $\hat{\rho}$ as $(\Gamma^{\mathcal{A}}_{\hat{\rho}})^{mn}_{ij}=\mathrm{Cov}(\hat{A}_i^{(m)},\hat{A}_j^{(m)})_{\hat{\rho}}$. Note that only the block-diagonal elements $(i=j)$ of this matrix appear on the right-hand side of Eq.~(\ref{eq.sepbound}). If the above covariance matrix is evaluated after replacing $\hat{\rho}$ with $\Pi(\hat{\rho})=\hat{\rho}_1\otimes\dots\otimes\hat{\rho}_N$ where $\hat{\rho}_i$ is the reduced density operator, obtained from $\hat{\rho}$ by tracing over all parties except the $i$th one, all of those inter-party correlations $(i\neq j)$ are removed, while the local terms $(i=j)$ remain unchanged. Hence, we arrive at the following expression for the local variances, $\sum_{i=1}^N\mathrm{Var}\left(\mathbf{c}_i\cdot\hat{\mathbf{A}}_i\right)_{\hat{\rho}} = \mathbf{c}^T \Gamma^{\mathcal{A}}_{\Pi(\hat{\rho})} \mathbf{c}$. Combining this with the expression for the quantum Fisher matrix, the separability criterion reads $W[\hat{\rho},\hat{A}(\mathbf{c})]=\mathbf{c}^T\left(Q^{\mathcal{A}}_{\hat{\rho}}-4\Gamma^{\mathcal{A}}_{\Pi(\hat{\rho})} \right)\mathbf{c}\leq 0$.
Since this condition must be satisfied for arbitrary vectors $\mathbf{c}$, it can be formulated independently of $\mathbf{c}$, as
\begin{align} \label{eq.boundrho}
Q^{\mathcal{A}}_{\hat{\rho}}-4\Gamma^{\mathcal{A}}_{\Pi(\hat{\rho})}\leq 0.
\end{align}
An entanglement witness is therefore found when the matrix $Q^{\mathcal{A}}_{\hat{\rho}}- 4\Gamma^{\mathcal{A}}_{\Pi(\hat{\rho})}$ has at least one positive eigenvalue. 
The criterion~(\ref{eq.boundrho}) can be equivalently stated as $\lambda_{\max}(Q^{\mathcal{A}}_{\hat{\rho}}- 4\Gamma^{\mathcal{A}}_{\Pi(\hat{\rho})})\leq 0$,
where $\lambda_{\max}(M)$ denotes the largest eigenvalue of the matrix $M$.

For pure states $\hat{\rho}=|\Psi\rangle\langle\Psi|$, the quantum Fisher matrix coincides, up to a factor of four, with the covariance matrix, i.e., $Q^{\mathcal{A}}_{|\Psi\rangle}=4\Gamma^{\mathcal{A}}_{|\Psi\rangle}$ \cite{BraunsteinPRL1994}. Thus, according to Eq.~(\ref{eq.boundrho}), every pure separable state must satisfy the condition
\begin{align}\label{eq.boundpure}
\Gamma^{\mathcal{A}}_{|\Psi\rangle}-\Gamma^{\mathcal{A}}_{\Pi(|\Psi\rangle)}\leq 0.
\end{align}
Conversely, if Eq.~(\ref{eq.boundpure}) is satisfied, then $\mathrm{Cov}(\hat{A}_i^{(m)},\hat{A}_j^{(n)})_{\ket{\Psi}}=0$, 
or equivalently $\langle\hat{A}_i^{(m)}\hat{A}_j^{(n)}\rangle_{|\Psi\rangle}=\langle\hat{A}_i^{(m)}\rangle_{|\Psi\rangle}\langle\hat{A}_j^{(n)}\rangle_{|\Psi\rangle}$ for all $i\neq j$ and all $n,m$ \cite{footnote1}.
If, additionally, each local set $\hat{A}_i^{(1)},\hat{A}_i^{(2)},\dots$ forms a complete set of observables, able to span the entire Hilbert space $\mathcal{H}_i$, for $i=1,...,N$, then this statement is only compatible 
with a product state, $|\Psi\rangle=|\varphi_1\rangle\otimes\dots\otimes|\varphi_N\rangle$. 
Hence, for each entangled pure state, a set of operators $\mathcal{A}$ can be found, such that the criterion~(\ref{eq.boundpure}) is violated. This means, the criterion~(\ref{eq.boundpure}) becomes necessary and sufficient for separability of pure states, 
while Eq.~(\ref{eq.boundrho}) is always a necessary criterion for arbitrary states.

\begin{figure}[tb]
\centering
\includegraphics[width=.48\textwidth]{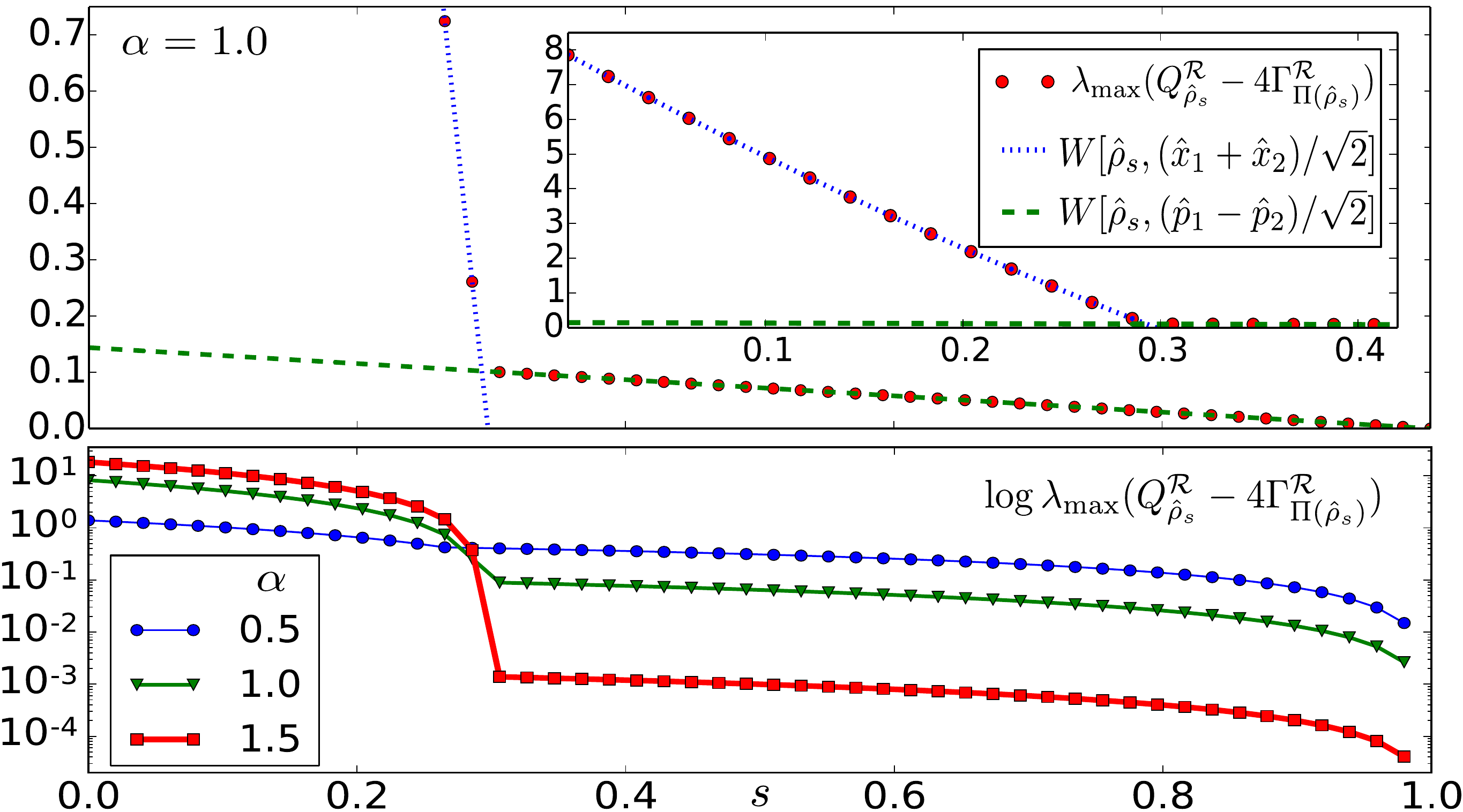}
\caption{\textit{Upper panel:} Detection of continuous-variable entanglement in the state~(\ref{eq.dephsup}) as a functon of $s$ for $\alpha=1$.
This state is undetected by the entropic separability criterion \cite{Walborn}. % for small values of $\alpha$.
For all values of $s$, a violation of Eq.~(\ref{eq.sepbound}) is observed when choosing $\hat{A}_1 = \hat{p}_1$ and $\hat{A}_2 = -\hat{p}_2$, as seen from the positivity of $W[\hat{\rho}_{s},(\hat{p}_1-\hat{p}_2)/\sqrt{2}]$ (green dashed line). 
For values of $s\lesssim 0.3$, an even stronger entanglement signature is observed for $\hat{A}_1 = \hat{x}_1$ and $\hat{A}_2 = \hat{x}_2$, as shown by $W[\hat{\rho}_{s},(\hat{x}_1+\hat{x}_2)/\sqrt{2}]$ (blue dotted) and highlighted in the inset. The optimized witness $\lambda_{\max}(Q^{\mathcal{R}}_{\hat{\rho}_s}-4\Gamma^{\mathcal{R}}_{\Pi(\hat{\rho}_s)})$ (red dots) confirms that, within the operators in $\mathcal{R}$, these choices of observables indeed yield the strongest possible violations of Eq.~(\ref{eq.sepbound}) for this state. \textit{Lower panel:} With increasing $\alpha$, the optimized witness gets stronger when $s\lesssim 0.3$ (position correlations dominate) and weaker for $s\gtrsim 0.3$ (momentum correlations dominate).}
\label{fig.dephasedsup}
\end{figure}

\textit{Application to continuous-variable entanglement.}---Let us first illustrate the applicability of the separability criterion derived here for the detection of mode-entanglement with continuous variables. A natural but arbitrary choice for the local observables $\hat{\mathbf{A}}_i$ are the phase-space operators $(\hat{x}_i,\hat{p}_i)^T$, such that the list of accessible observables $\mathcal{A}$ is given by $\mathcal{R}=\{\hat{x}_1,\hat{p}_1,\hat{x}_2,\hat{p}_2,\dots,\hat{x}_N,\hat{p}_N\}$ (we henceforth drop the vector notation for clarity). Gaussian states are fully characterized in terms of their covariance matrix $\Gamma^{\mathcal{R}}_{\hat{\rho}}$ \cite{Braunstein}, and their bipartite entanglement is efficiently captured by separability criteria based on Heisenberg's uncertainty relation \cite{Duan,Giovannetti}. The strongest criterion of this kind follows from the sharpest uncertainty relation, which is formulated in terms of entropic quantities \cite{Walborn}. To test the limits of these criteria, they are applied to non-Gaussian states, such as a partially dephased superposition state of the form
\begin{align}\label{eq.dephsup}
\hat{\rho}_s=N(\alpha,s)&\big[|\alpha,\alpha\rangle\langle \alpha,\alpha|+|-\alpha,-\alpha\rangle\langle -\alpha,-\alpha|\\
&+(1-s)\left(|-\alpha,-\alpha\rangle\langle \alpha,\alpha|+|\alpha,\alpha\rangle\langle -\alpha,-\alpha|\right)\big]\notag,
%\hat{\rho}_s=N(\alpha,s)&\big[|\alpha,-\alpha\rangle\langle \alpha,-\alpha|+|-\alpha,\alpha\rangle\langle -\alpha,\alpha|\\
%&+(1-s)\left(|-\alpha,-\alpha\rangle\langle \alpha,\alpha|+|\alpha,\alpha\rangle\langle -\alpha,-\alpha|\right)\big]\notag,
\end{align}
where $|\alpha\rangle$ is a coherent state, $0\leq s\leq 1$ is a parameter and $N(\alpha,s)$ a normalization constant. 
The entropic separability criterion, and with it all other uncertainty-based criteria \cite{Duan,Simon,Giovannetti}, are unable to detect the entanglement of $\hat{\rho}_s$ for small values of $\alpha$ \cite{Walborn}. 
The criterion~(\ref{eq.boundrho})  --  using only the local operators contained in $\mathcal{R}$ -- detects the entanglement of the state $\hat{\rho}_s$ for all values of $s$ and $\alpha$, except at $s=1$ where the state is separable. This is illustrated in Fig.~\ref{fig.dephasedsup} for values of $\alpha$ in the undetected region of the entropic criterion.

The position and momentum observables contained in $\mathcal{R}$ represent one of many possible sets of local operators that can be used to construct a  separability criterion from Eq.~(\ref{eq.sepbound}) in a continuous-variable system. Another choice of local observables is given by the local number operators $\hat{n}_i$. In quantum optical experiments, the local number fluctuations are accessible in a variety of platforms \cite{Bakr10,Schindler,Lucke}, and comparison with the quantum state's sensitivity to a collective phase shift $\exp(i\theta\hat{N})$, generated by $\hat{N}=\sum_{i=1}^N\hat{n}_i$ leads to the experimentally convenient separability criterion $F_Q[\hat{\rho}_{\mathrm{sep}},\hat{N}]\leq 4\sum_{i=1}^N\mathrm{Var}(\hat{n}_i)_{\hat{\rho}_{\mathrm{sep}}}$.

The various criteria obtained from Eq.~(\ref{eq.sepbound}) for different choices of local operators may also be combined 
to generate bounds whose verification does not require direct measurements of the local variances. Consider for example a continuous variable system of $N$ parties (modes), for which $F_Q[\hat{\rho}_{\mathrm{sep}},\hat{X}]$ and $F_Q[\hat{\rho}_{\mathrm{sep}},\hat{P}]$ have been independently probed, with $\hat{P}=\sum_{i=1}^N\hat{p}_i$ and $\hat{X}=\sum_{i=1}^N\hat{x}_i$. The sum of the two corresponding inequalities~(\ref{eq.sepbound}) then yields the criterion
\begin{align}\label{eq.combinedbound}
F_Q[\hat{\rho}_{\mathrm{sep}},\hat{X}]+F_Q[\hat{\rho}_{\mathrm{sep}},\hat{P}]&\leq 4\sum_{i=1}^N\left(\mathrm{Var}(\hat{x}_i)+\mathrm{Var}(\hat{p}_i)\right)_{\hat{\rho}_{\mathrm{sep}}}\notag\\
&\leq 4\sum_{i=1}^N\left(\langle\hat{x}_i^2\rangle_{\hat{\rho}_{\mathrm{sep}}}+\langle\hat{p}_i^2\rangle_{\hat{\rho}_{\mathrm{sep}}}\right)\notag\\
&= 4\sum_{i=1}^N(2n_i+1)=4(2n+N),
\end{align}
where $n_i=\langle\hat{n}_i\rangle_{\hat{\rho}}$ is the average particle number in mode $i$ with $\hat{n}_i=\hat{a}^{\dagger}_i\hat{a}_i$ and $\hat{a}_j=(\hat{x}_j+i\hat{p}_j)/\sqrt{2}$. The second inequality is saturated if and only if $\langle \hat{x}_i\rangle_{\hat{\rho}_{\mathrm{sep}}}=\langle \hat{p}_i\rangle_{\hat{\rho}_{\mathrm{sep}}}=0$, for all $i=1,\dots,N$. If the number of modes $N$ and the total number of particles $n=\sum_{i=1}^Nn_i$ are known from independent measurements, Eq.~(\ref{eq.combinedbound}) can be used as an entanglement witness without measurement of the local variances.

\begin{figure}[tb]
\centering
\includegraphics[width=.48\textwidth]{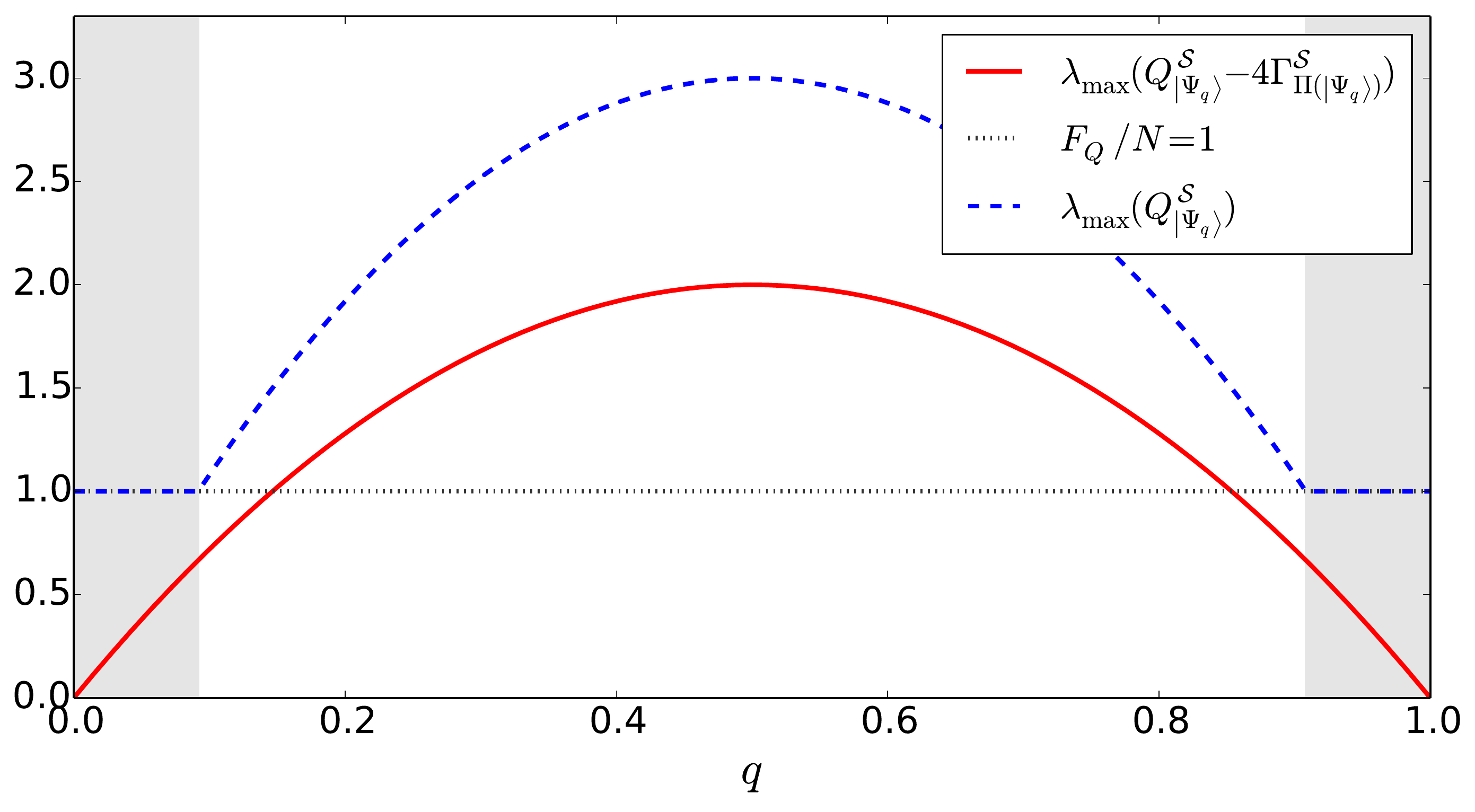}
\caption{
Detection of qubit entanglement in the state Eq.~(\ref{eq.shotnoiseent}) as a functon of $q$ and for $N=3$.
This state is undetected by the shot-noise limit in the gray shaded area: the optimized quantum Fisher information (blue dashed) does not exceed the separable limit of $1$, which corresponds to the shot-noise limit of $F_Q=N$ (gray dashed line). For $N=3$ the undetected region is the largest.
In contrast, the red solid line shows $\lambda_{\max}(Q^{\mathcal{S}}_{|\Psi_q\rangle} - 4\Gamma^{\mathcal{S}}_{\Pi(|\Psi_q\rangle)})$, 
which is positive and thus violates the separability criterion~(\ref{eq.boundpure}) for all values of $0 < q < 1$. For $q=0,1$ the state is separable.}
\label{fig.shotnoiseentanglement}
\end{figure}

\textit{Application to discrete-variable entanglement.}---From Eq.~(\ref{eq.sepbound}) it is possible to derive state-independent upper bounds, using $4 \mathrm{Var}(\hat{A})_{\hat{\rho}}\leq (\lambda_{\max}(\hat{A})-\lambda_{\min}(\hat{A}))^2$, which holds for all $\hat{\rho}$ and $\lambda_{\min/\max}(\hat{A})$ denote minimal and maximal eigenvalues of $\hat{A}$. This yields the state-independent separability bound
\begin{align}\label{eq.stateindep}
F_Q\bigg[\hat{\rho}_{\rm sep},\hat{A}(\mathbf{c})\bigg]&\leq \sum_{i=1}^N\left(\lambda_{\max}(\mathbf{c}_i\cdot\hat{\mathbf{A}}_i)-\lambda_{\min}(\mathbf{c}_i\cdot\hat{\mathbf{A}}_i)\right)^2.
%\notag\\&\leq \sum_{i=1}^N\left(\sum_m\left[\lambda_{\max}(c^{(i)}_m\hat{A}^{(i)}_m)-\lambda_{\min}(c^{(i)}_m\hat{A}^{(i)}_m)\right]\right)^2\notag\\&= \sum_{i=1}^N\left(\sum_m |c^{(i)}_m|\left[\lambda_{\max}(\hat{A}^{(i)}_m)-\lambda_{\min}(\hat{A}^{(i)}_m)\right]\right)^2\notag\\
%&\leq \sum_{i=1}^N\left(\sum_m |c^{(i)}_m|^2\sum_n\left[\lambda_{\max}(\hat{A}^{(i)}_n)-\lambda_{\min}(\hat{A}^{(i)}_n)\right]^2\right)\notag\\
%&= \sum_{i=1}^N\left(|\textbf{c}^{(i)}|^2\sum_n\left[\lambda_{\max}(\hat{A}^{(i)}_n)-\lambda_{\min}(\hat{A}^{(i)}_n)\right]^2\right)
\end{align}
In the special case of $N$ qubit systems, we recover the shot-noise bound $F_Q[\hat{\rho}_{\rm sep},\sum_{i=1}^N \mathbf{c}_i\cdot\hat{\boldsymbol{\sigma}}_i/2]\leq N$, 
whose violation identifies entangled states of $N$ qubits that are useful for sub-shot-noise interferometry \cite{PS09} when $\mathbf{c}_i=\mathbf{n}\in\mathbb{R}^3$ is a unit vector (leading to $|\mathbf{c}|^2=N$). Here, $\hat{\boldsymbol{\sigma}}_i=(\hat{\sigma}_i^{(x)},\hat{\sigma}^{(y)}_i,\hat{\sigma}^{(z)}_i)$ is the vector of Pauli matrices for the $i$th qubit.
Certain entangled $N$-qubit states, however, cannot be detected by the shot-noise criterion, even if the state is further optimized by means of local unitary manipulations \cite{Hyllus}, i.e., by optimization of the $\mathbf{c}_i$ under the normalization constraints $|\mathbf{c}_i|^2=1$. More generally, without respecting these constraints, we obtain the separability criterion $\lambda_{\max}(Q^{\mathcal{S}}_{\hat{\rho}_{\rm sep}})\leq 1$, where $\mathcal{S}=\{\hat{\boldsymbol{\sigma}}_1/2,\dots,\hat{\boldsymbol{\sigma}}_N/2\}$. Yet, the entanglement of states of the form
\begin{align}\label{eq.shotnoiseent}
|\Psi_q\rangle=\sqrt{q}|0\rangle^{\otimes N}+\sqrt{1-q}e^{i\varphi}|1\rangle^{\otimes N},
\end{align}
will be overlooked by any of these state-independent bounds when $q\leq(1-\sqrt{(N-1)/N})/2$ or $q\geq (1+\sqrt{(N-1)/N})/2$ and $N\geq 3$ \cite{Hyllus}. In contrast, the stronger state-dependent criterion~(\ref{eq.boundrho}), which for pure states reduces to~(\ref{eq.boundpure}), is necessary and sufficient for all pure states, since the Pauli matrices span a complete set of qubit observables---together with the identity operator which commutes with all operators and can therefore be omitted. Figure~\ref{fig.shotnoiseentanglement} shows that, indeed, the stronger criterion~(\ref{eq.boundrho}) (red continuous line) detects the entanglement of $|\Psi_q\rangle$ for arbitrary values of $0< q< 1$, while the optimized quantum Fisher information (blue dashed line) does not overcome the shot-noise bound (gray dotted line) in the intervals mentioned above (gray shaded areas). Choosing an orientation along the $z$-axis, i.e., $\mathbf{c}_i=(0,0,1)^T$ for all $i=1,\dots,N$, yields the largest positive values of the witness $W$ [Eq.~(\ref{eq.W})], for all parameters $q$, while it maximizes the quantum Fisher information only outside of the gray-shaded parameter regime.

\textit{Entanglement depth, hybrid variables and experiments.}---The state-independent bounds~(\ref{eq.stateindep}) can be generalized to distinguish among the hierarchical classes of $k$-partite entanglement \cite{FImulti}. Consider an $N$-mode state which contains at most $k$-partite entanglement, i.e., $\hat{\rho}_{k\rm-prod}=\sum_\gamma p_\gamma |\varphi^\gamma_1\rangle\langle\varphi^\gamma_1|\otimes|\varphi^\gamma_2\rangle\langle\varphi^\gamma_2|\otimes\cdots$, where each of the states $|\varphi^\gamma_l\rangle$ describes $N^{(\gamma)}_l\leq k$ modes with $\sum_lN^{(\gamma)}_l=N$ for all $\gamma$. An upper bound for the quantum Fisher information is given by
\begin{align}\label{eq.kpartite}
F_Q[\hat{\rho}_{k\rm-prod},\hat{A}(\mathbf{c})]
\leq \Delta_{\max}^2(sk^2+r^2),
\end{align}
where $s=\lfloor N/k\rfloor$ and $r=N-sk$ and the maximum spectral span of all local operators is denoted by $\Delta_{\max}=\max_i\{\lambda_{\max}(\mathbf{c}_i\cdot\hat{\mathbf{A}}_i)-\lambda_{\min}(\mathbf{c}_i\cdot\hat{\mathbf{A}}_i)\}$. The above result is obtained following Refs.~\cite{FImulti} together with $\lambda_{\max}(\sum_i\hat{A}_i)\geq \sum_i\lambda_{\max}(\hat{A}_i)$ and $\lambda_{\min}(\sum_i\hat{A}_i)\leq \sum_i\lambda_{\min}(\hat{A}_i)$, which is ensured by Weyl's inequality \cite{Weyl}, and thereby generalizes the $N$-qubit result of Refs.~\cite{FImulti} to the case of unequal, arbitrary subsystems. Hence, whenever the spectrum of the local operators is bounded, Eq.~(\ref{eq.kpartite}) provides a criterion not only to test if any entanglement is present, but also how many of the $N$ modes are entangled. Besides finite-dimensional systems \cite{FImulti,Varenna}, there may also be applications to continuous variables if further knowledge about the system limits the spectral range $\Delta_{\max}$. For example if a gas is contained in a trap of extension $L$, the spectral span of the position operator cannot exceed $L$. In such a system, any observation of $F_Q[\hat{\rho},\hat{X}]>L^2(sk^2+r^2)$ indicates entanglement of $k$ modes.

Entanglement detection protocols have traditionally been developed for homogeneous systems of either discrete or continuous variables \cite{Braunstein,GuehneToth}. Nevertheless, hybrid correlations between the two are generated in many different experiments \cite{Haroche,Wineland,Bellini,Laurat,Berkeley14,Hefei15}, and their potential for quantum information processing is recognized \cite{Andersen}. One of the advantages of the separability criterion~(\ref{eq.sepbound}) is its independence of the Hilbert space structure and dimension, allowing for the possibility of witnessing hybrid entanglement. As a simple example, consider the composition of a two-level atom, coupled to a single harmonic oscillator mode \cite{Haroche,Wineland}. Correlated states such as $|\phi_n\rangle=(|0,n\rangle+|1,n+1\rangle)/\sqrt{2}$, where $|n\rangle$ denotes a Fock state of $n$ excitations, are produced in ion-trap experiments \cite{Berkeley14}. A suitably designed hybrid criterion such as $F_Q[\hat{\rho}_{\mathrm{sep}},\hat{\sigma}^{(x)}+\hat{x}]\leq 4\mathrm{Var}(\hat{\sigma}^{(x)})_{\hat{\rho}_{\mathrm{sep}}}+4\mathrm{Var}(\hat{x})_{\hat{\rho}_{\mathrm{sep}}}$ is able to reveal this entanglement. Recall from Eq.~(\ref{eq.boundpure}) that for pure states, separability requires the absence of inter-party covariances. The entanglement of the state $|\phi_n\rangle$ can thus be attributed to the coherences that lead to $\langle \sigma^{(x)}\hat{x}\rangle_{|\phi_n\rangle}\neq \langle\sigma^{(x)}\rangle_{|\phi_n\rangle}\langle \hat{x}\rangle_{|\phi_n\rangle}$, and ultimately cause the violation of the separability criterion above.

Before we conclude, let us briefly discuss the experimental implementation of our proposed entanglement criteria. In order to measure the witness~(\ref{eq.W}), two quantities need to be obtained. On the one hand, the variances of the local operators $\mathbf{c}_i\cdot\hat{\mathbf{A}}_i$ need to be measured. Single-site \textit{addressing} is not needed to achieve this. Instead, only the much less demanding resolved imaging of the individual constituents is required. Such measurements are possible in current experiments with, e.g., multi-mode photonic states \cite{Fabre,Bellini,Laurat,Treps}, trapped ions \cite{Monz,Schindler}, as well as with cold atoms distinguished by multi-well potentials or internal states \cite{Strobel,Lucke,Peise,BECfootnote}, and under quantum-gas microscopes \cite{Bakr10,Bloch2011,FermiMicroscope1,FermiMicroscope2,FermiMicroscope3,Greiner}. On the other hand, the Fisher information can be extracted, e.g., following the method of Ref.~\cite{Strobel} by determining the impact of a collective unitary operation with no need for local measurements, see also \cite{Bohnet,Monroe,PezzePRE,hauke,frowis}. Measurements of the Fisher information can be completely avoided by using the lower bound $F_Q[\hat{\rho},\hat{A}]\geq |\langle [\hat{A},\hat{B}]\rangle_{\hat{\rho}}|^2/\mathrm{Var}(\hat{B})_{\hat{\rho}}$, which holds for arbitrary operators $\hat{A}$ and $\hat{B}$ \cite{Varenna}. Together with the separability condition $W[\hat{\rho},\hat{A}(\mathbf{c})]\geq 0$ from Eq.~(\ref{eq.W}), we obtain the simple criterion,
\begin{align}
\mathrm{Var}(\hat{A}(\mathbf{c}))_{\Pi(\hat{\rho})}\mathrm{Var}(\hat{B})_{\hat{\rho}}\geq \frac{|\langle [\hat{A}(\mathbf{c}),\hat{B}]\rangle_{\hat{\rho}}|^2}{4},
\end{align}
whose violation indicates entanglement of $\hat{\rho}$.

\textit{Conclusions.}---We have introduced a family of entanglement criteria that are applicable to multipartite systems of discrete and/or continuous variables. The criteria are based on a comparison of the Fisher information, which expresses the sensitivity of the quantum state to a collective unitary transformation, with the sum of variances of the local operators that generate this transformation. We have illustrated the applicability with examples of spins and continuous variables, showing in both cases that our criteria are able to detect the entanglement of states that remain undetected with commonly employed methods that defined the state of the art throughout the last years. In particular, we have constructed entanglement criteria that are necessary and sufficient for all pure states. Since any set of accessible local operators can be used to generate a sufficient entanglement criterion, the strategy presented here allows for versatile adaptations to a variety of experiments, and can be readily implemented within the current technology.
%Measurements of the Fisher information have been experimentally demonstrated with Bose-Einstein condensates \cite{Strobel} and trapped ions \cite{Bohnet,Monroe} using a platform-independent method. Local measurements of correlated degrees of freedom are required to obtain the local variances. These are available in a variety of quantum systems, including trapped ions \cite{Monz}, photons \cite{Bellini,Laurat,Treps} and ultra-cold atoms \cite{Lucke,Bakr10, Bloch2011,Peise,BECfootnote}. 
%Hence, we expect the results provided here to improve entanglement detection methods in present-day experiments.


\begin{thebibliography}{50}
\bibitem{FIfootnote}
To measure the dependence of $\hat{\rho}(\theta)$ on $\theta$, a probability distribution $p(m|\theta)=\mathrm{Tr}[\hat{\rho}(\theta)\hat{P}_{m}]$ is recorded, where $\hat{M}$ is some observable with eigenvalues $m$ and associated projectors $\hat{P}_{m}$. The Fisher information is defined as $F_{\hat{M}}[ \hat{\rho}, \sum_{i=1}^N \hat{A}^{(i)}]=\sum_{m}\frac{1}{p(m|\theta)}\left(\frac{\partial p(m|\theta)}{\partial \theta}\right)^2$. Its value may depend on $\theta$ but since the bound~(\ref{eq.sepbound}) holds for all values of $\theta$ we do not denote this dependence explicitly.
\bibitem{footnote1}
The matrices $\Gamma^{\mathcal{A}}_{|\Psi\rangle}$ and $\Gamma^{\mathcal{A}}_{\Pi(|\Psi\rangle)}$ have, by construction, the same block diagonal terms and, 
in particular, the same trace.
This implies that the eigenvalues $\lambda_i$ of the matrix $\Delta\Gamma^{\mathcal{A}}_{|\Psi\rangle} =  \Gamma^{\mathcal{A}}_{|\Psi\rangle}-\Gamma^{\mathcal{A}}_{\Pi(|\Psi\rangle)}$
satisfy $\sum_i\lambda_i = \mathrm{Tr}[\Delta\Gamma^{\mathcal{A}}_{|\Psi\rangle}] = 0$.
This is only compatible with Eq.~(\ref{eq.boundpure}) if all the eigenvalues of $\Delta\Gamma^{\mathcal{A}}_{|\Psi\rangle}$ are equal to zero, i.e. 
$\Gamma^{\mathcal{A}}_{|\Psi\rangle} = \Gamma^{\mathcal{A}}_{\Pi(|\Psi\rangle)}$.
\bibitem{BECfootnote} 
We note that when demonstrating particle entanglement among indistinguishable particles, as in the case, for instance, of 
\cite{Lucke, Strobel}, local observables cannot be measured. 
However, local measurements of the modes (encoded into different hyperfine levels or trapping wells) of a Bose-Einstein condensate
can be used to study mode entanglement based on Eq.~(\ref{eq.sepbound}).
\end{thebibliography}
\end{document}